\documentclass[12pt]{article}
\usepackage{graphicx}
\usepackage{amsmath,amssymb,amsfonts,multirow}
\usepackage{mathrsfs}
\usepackage{latexsym}
\usepackage{bm}
\usepackage{bbm}
\usepackage{cite}
\usepackage{subfigure}
\usepackage{mathrsfs}
\usepackage{indentfirst}
\usepackage{amsmath}
\usepackage{amssymb}
\usepackage{cases}
\usepackage{mathrsfs}
\usepackage{setspace}
\usepackage{threeparttable}
\usepackage{booktabs}

\usepackage[colorlinks]{hyperref}
\hypersetup{citecolor=blue,linkcolor=blue,urlcolor=blue}

\begin{document}

\title{\Large\bf Pyramid scheme in stock market: a kind of financial market simulation}
\vspace{0.2truecm}

\author{Yong Shi$^{1,2,3}$,\ Bo Li$^{1,2,3}$\footnote{Email:libo312@mails.ucas.ac.cn} \  and Guangle Du$^{4,5}$
\\
\\
  \textit{$^{\small 1}$\small School of Economics and Management,}\\
  \textit{\small University of Chinese Academy of Sciences, Beijing 100190, China}\\
  \textit{$^{\small 2}$\small Research Center on Fictitious Economy and Data Science},\\
  \textit{\small Chinese Academy of Sciences, Beijing 100190, China}\\    
  \textit{$^{\small 3}$\small Key Laboratory of Big Data Mining and Knowledge Management,}\\
  \textit{\small Chinese Academy of Sciences, Beijing 100190, China}\\
  \textit{$^{\small 4}$\small Wenzhou Institute, University of Chinese Academy of Sciences, Wenzhou, Zhejiang 325001, China} \\
  \textit{$^{\small 5}$\small School of Physical Sciences, University of Chinese Academy of Sciences, Beijing 100049, China} 
}
\date{}
\maketitle
\begin{abstract}
Artificial stock market simulation based on agent is an important means to study financial market. Based on the assumption that the investors are composed of a main fund, small trend and contrarian investors characterized by four parameters, we simulate and research a kind of financial phenomenon with the characteristics of pyramid schemes. Our simulation results and theoretical analysis reveal the relationships between the rate of return of the main fund and the proportion of the trend investors in all small investors, the small investors' parameters of taking profit and stopping loss, the order size of the main fund and the strategies adopted by the main fund. Our work are helpful to explain the financial phenomenon with the characteristics of pyramid schemes in financial markets, design trading rules for regulators and develop trading strategies for investors.
\end{abstract}

\newpage

\section{Introduction}
Simulation of artificial financial market based on multi-agent models is an important method to search for the dynamic laws in financial market~\cite{chiarella2006asset, georges2009learning, rekik2014agent, brock2005evolutionary}, study the statistical properties of financial time series~\cite{lux1999scaling, mota2017detailed, PhysRevE.60.1390, gopikrishnan2000scaling}, and evaluate the possible effects of the financial regulations and/or rules~\cite{2012Agent, mizuta2016brief}. The multi-agent models for researching the dynamics in financial market usually assume that the agents are heterogeneous, then study the behaviors of the agents such as herd behavior or contrarian behavior~\cite{zhao2011herd, liang2013contrarian, kononovicius2019order}, and the influences of the factors such as event shocks on market dynamics~\cite{chiarella2006asset, georges2009learning, rekik2014agent, brock2005evolutionary}. The universal statistical properties in financial time series, autocorrelation and scaling exists in returns, volatility clustering, etc., generically called stylized facts~\cite{mantegna1994stochastic, gopikrishnan2000scaling, PhysRevE.60.1390}, can be reproduced by artificial financial market simulation~\cite{lux1999scaling, mota2017detailed}. Artificial market studies for evaluating the effects of the financial regulations and/or rules can give insights to the discussion on whether price variation limits and short selling regulation can prevent bubbles and crushes or not, and estimate the effects of tick size, usage rate of dark pools, rules for investment diversification, order matching systems' speed on financial exchanges~\cite{2012Agent, mizuta2016brief}.

In addition to the research objectives above, we can also roughly divide artificial financial market models from the characteristics of theirselves into the agent models developed from minority games\cite{challet2000modeling, challet2001stylized, johnson2000trader, jefferies2001market}, the agent models based on existing physical models~\cite{stauffer1998can, hong2014multiscale, yu2012lattice}, the agent models based on order book and so on~\cite{preis2006multi, mota2017detailed}.
Recently, some studies have combined deep learning with artificial financial markets, Ref.~\cite{raman2019financial} considers the case that there exists a class of agents who utilize deep learning to make decisions, and Ref.~\cite{maeda2020deep} involves taking advantage of deep reinforcement learning in agent based financial market simulation to learn a robust investment strategy with an attractive risk-return profile.

In this work, we propose an agent model to simulate and research a kind of financial phenomenon with the characteristics of the pyramid schemes~\cite{gastwirth1977probability, gastwirth1984two, feng2020case, shi2019pyramid}, which have organizers who induce people to participate in the schemes with high returns, and the participants also recruit the next generation of participants until the schemes can not continue. In the secondary market of a certain security, there is often a financial phenomenon with the characteristics of the pyramid schemes: the main funds with a large amount of money and securities may act as the organizers, who influence the price of the security through big orders to induce ordinary trend investors to follow the trend one after another, and profit through reverse operation finally.

The financial phenomenon with the characteristics of pyramid schemes above is somewhat similar to herd behavior in financial markets~\cite{zhao2011herd, cipriani2009herd, bikhchandani2000herd, avery1998multidimensional}, they are all involved in the behavior of following the trend, and there are some works about agent based models with herding or contrarian behaviors, for instances, Zhao et al.~\cite{zhao2011herd} and Liang et al.~\cite{liang2013contrarian} deeply study the mechanisms of herding behaviors and contrarian behaviors in adaptive systems.
But the difference between them is obvious, the core of the financial phenomenon with the characteristics of pyramid schemes is that the main fund inducing the small investors to follow the trend and making profits, while the herd behavior market emphasizes the herd mentality of the ordinary investors in financial markets.
In order to study the financial phenomenon with the characteristics of pyramid schemes, we simulate this kind of financial market by building an agent model with a special investor structure. 
While the difference between the two phenomena is obvious, the core of the financial phenomenon with the characteristics of pyramid schemes is the main fund inducing the small investors to follow the trend and making profits, while the herd behavior emphasizes the herd mentality of the ordinary investors in financial markets.
In order to study the financial phenomenon with the characteristics of pyramid schemes, we simulate this kind of financial market by building an agent model with a special investor structure.
In our proposed model, we assume there is only one main fund in the market, and the small investors include homogeneous trend and contrarian investors described by 4 parameters as follows: $Type$, which represents the type of the strategy an investor used which is depended on his or her specific trading directions relative to the market trend, trend or contrarian, and is different from many existing works on financial market simulation which often divide investors into fundamental investors, technical investors and other investors\cite{lux1999scaling, mota2017detailed, challet2000modeling, challet2001stylized, raberto2001agent}; $r_{market}$, which determines when a investor decides to participate in trading according to the change of the security price in a certain trading period, due to the confidentiality of investors' strategies in the real market, we don't specify the mechanisms of the investors' strategies as in Refs.~\cite{chiarella2006asset, mota2017detailed, challet2000modeling, challet2001stylized, jefferies2001market}, and only assume that the parameter $r_{market}$ of each investor is different and make some assumptions about the parameters' overall distribution; $r_{profit}$ and $r_{loss}$, which denote the take-profit point and stop-loss parameters set by investors. In the simulations, in order to fit the real market better, we also set the activation probability $p_{active}$ for each small investor as in Ref.~\cite{mota2017detailed}.

Based on the artificial financial market established, we make use of computer simulation and theoretical analysis to reveal the relationships between the rate of return of the main fund and the proportion of the trend investors in all small investors, the small investors' parameters of taking profit and stopping loss, the order size of the main fund and the strategies adopted by the main fund. Our work clearly explains the financial phenomenon with the characteristics of pyramid schemes in financial markets, and it can help regulators to design trading rules and assist investors in developing trading strategies.

The rest of the paper is organized as follows: In the next section, we introduce our proposed agent model. In Sec.~\ref{simulations} the simulation results and preliminary analysis are displayed, and Sec.~\ref{Theoretical_analysis} gives a further analysis. Some discussions and conclusions are given in Sec.~\ref{conclusion}.

\section{Model}

\subsection{General aspects}
As the objective of our proposed model is to study the financial phenomenon with the characteristics of pyramid schemes, we assume that there are three types of agents in our models, the first one is the main fund, and the other two are small contrarian or trend investors who adopt contrarian strategies or trend strategies to trade. In the trading process, the main fund buy or sell large orders to induce small trend investors, and then profit through reverse operation. The expected rate of return of the main fund largely determines its behavior, which plays a leading role in the financial phenomenon we research, thus our object is to research the relationships between the profit level of the main fund and the strategies of the main fund adopts, the order size of the main fund and the structure of the small investors. The characteristics of the different agents and specific trading rules are described in detail below.

\subsection{Types of agents}
The main fund in our model is the agent who has a large amount of money or the traded security relative to other agents, and as a result, its trading behaviors can dominate the change of the security price and it profits from this advantage. It should be noted that the artificial financial market we simulate only trades one kind of security. In order to simplify the analysis process and better study the problems we are concerned about, there is only one main fund in our model, and it does not have the ability to completely manipulate the market, that is to say, the main fund has no ability to swallow all the buy or sell orders from the small investors.

The other two types of agents are small trend and contrarian investors, who employ trend strategies and contrarian strategies respectively in trading. Trend agents participate in trading according to the trend of the security price, if they think there is a sign of an upward trend in the security price, they go long the security, otherwise they go short the security. Contrarian agents go short when the security price rises more than a certain extent, and go long when the security price falls by more than a certain range.
We do not divide investors into fundamental investors and technical analysis investors as Refs.~\cite{lux1999scaling, mota2017detailed} do, where the fundamental investors and the technical analysis investors correspond to the investors using contrarian strategies and the investors using trend strategies respectively. But in fact, fundamental investors will also follow the upward trend as the valuation of the security increases, and technical analysis investors often adopt contrarian strategies. Therefore, we adopt a more reasonable way to distinguish investors based on their investment strategies rather than on whether they adopt technical analysis or fundamental analysis.

For the investors with trend strategies and contrarian strategies in our model, the investors with trend strategy trade at the market price and the investors with contrarian strategies trade at the limit price, we use parameter combinations $(Type,$  $r_{market},$ $r_{profit},$ $r_{loss})$ to characterize them, where $Type$ represents the types of strategies used by investors, trend or contrarian; $r_{market}$ refers to the trigger signals for trend investors or determines the prices of limit orders for contrarian investors, and the trend investors will participate in the trading when the security price's range of rise and fall is more than $r_{market}$ over a certain period of time, while the contrarian investors place limit orders at $(1+r_{market})P_0$, where $P_0$ is the initial price of the security in the period of time; $r_{profit}$ and $r_{loss}$ denote the take-profit and stop-loss parameters set by investors respectively.

Generally, different parameters $r_{profit}$ and $r_{loss}$ can describe investors with different behavioral characteristics. For instance, for an investor with disposition effect~\cite{weber1998disposition}, the absolute value of stop-loss parameter is greater than that of take-profit parameter.

In real financial markets, due to the confidentiality of investment strategies which determine $r_{market}$, we do not know investors' specific investment strategies and when and how they will participate in the trading. However, certain prices of securities may gather a large number of investors to participate in trading for some reasons, for instances, when some common technical indicators such as moving averages (MAs) indicate buy signals, or the prices of securities reach key positions such integer positions, a large number of investors will participate in trading, so we can make a hypothesis that the parameters $r_{market}$ are concentrated in the vicinity of some values and are normally distributed. Fig.~1 shows the schematic diagram reflecting the structure of small investors, the parameters $r_{market}^1$, $r_{market}^2$, $r_{market}^3$, $r_{market}^4$ and $r_{market}^5$ denote the rate of increase and decrease relative to the initial price in the figure are different triggering signals for the trend strategies and contrarian investors, and these parameters are normally distributed.

\begin{figure}
	\centering
	\includegraphics[width=1\textwidth]{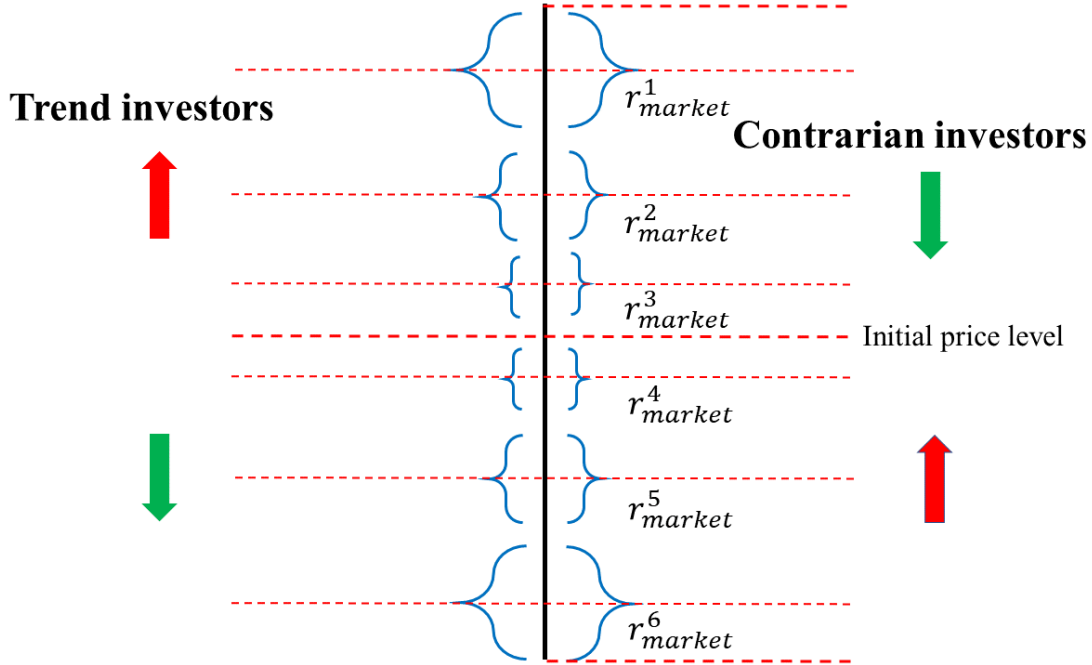}
	\caption{The schematic diagram of the trend investors and contrarian investors in our proposed model.}
	\label{Schematic}
\end{figure}

Small investors in our model do not have the ability to affect the security price, they are passive to participate in trading, only the main fund has the ability to influence the security price. In addition, each small investor only sells or buys orders with size 1, but this does not affect the representativeness of our model, because in actual financial transactions, an small investor who can buy and sell orders with the size greater than 1 can be regarded as a collection of investors who can only buy and sell orders with size 1.

\subsection{Details of trading rules and process}

In our model, the orders from the main fund and the small investors are divided into market orders which are active buy or sell orders, and limit orders which are passive.
The matchmaking trading mode employed in our model is the fashion of time priority and price priority, and we define the process from the issuance of a market order of the main fund, to the time time when there is no new transaction unless the main fund issues a new market order as a trading period. In the following, we describe the procedures of the first trading period in a step-by-step way from the issuance of a market buy order of the main fund to the end of all transactions, and any other trading periods are similar. The procedures are:

(\romannumeral1) The main fund issues a market buy order with the size $N_{mf}$ ($N_{mf}$ is a positive integer), and it concludes transactions with the limit sell orders placed by $N_{mf}$ contrarian investors with $r_{market}>0$. After trading, the market price of the security will rise. It should be noted here that we set the activation probability $p_{active}$ for each contrarian investor, so even if its $r_{market}$ satisfies the trading conditions, it only has the probability of $p_{active}$ to trade. The same is true for trend investors.

(\romannumeral2) The contrarian investors who trades with the main fund in procedure (\romannumeral1) set take-profit and stop-loss orders according to their parameters $r_{profit}$ and $r_{loss}$, and the take-profit orders are limit orders while the stop-loss orders are market orders. In our setting, the small investors who have taken profit or stopped loss will no longer participate in trading in the same trading period.

(\romannumeral3) The rise of the security price caused by the buying of the main fund triggers the trend traders whose parameters $r_{market}$ are no more than the rise to buy. The triggered trend traders will cause the price to rise again.

(\romannumeral4) The trend and contrarian investors trade in procedure (\romannumeral3) set take-profit and stop-loss orders according to their parameters $r_{profit}$ and $r_{loss}$. Same as contrarian investors, the take-profit orders of the trend investors are limit orders and the stop-loss orders are market orders.

(\romannumeral5) The rise caused by the triggered trend traders may trigger new trend traders to buy. This process will continue until there are no trend investors are triggered, and the trading period ends when the process stops.

Tab.~\ref{structure} shows parameters and numbers of trend investors and contrarian investors, where $Mean(r_{market})$ and $\sigma(r_{market})$ in each row are the average and standard deviation of the normal distribution satisfied by the parameter $r_{market}$ of the investors in the same row, and the parameter $ratio$ denotes the ratio of the trend investors to the contrarian investors. It is worth noting that there are 20000 contrarian investors with the parameter $r_{market}$ of 0.1 and -0.1 respectively, but no trend investors with the parameter $r_{market}$ of 0.1 and -0.1. The purpose of our design is to be close to the real market, because when the price of a security deviates greatly from its due value, it will attract a large number of contrarian investors to participate in the trading.

\begin{table}[tp]
	\centering
	\fontsize{8}{10}\selectfont
	\begin{threeparttable}
		\caption{Parameters and numbers of trend investors and contrarian investors.}
		\label{structure}
		\begin{tabular}{ccccccccc}
			\toprule
			%\multirow{2}{*}{Model}&
			\multicolumn{4}{c}{Trend investors}&
			\multicolumn{4}{c}{Contrarian investors}\cr
			\cmidrule(lr){1-4} \cmidrule(lr){5-8}
			$Mean(r_{market})$&$\sigma(r_{market})$&Number&$p_{active}$
			&$Mean(r_{market})$&$\sigma(r_{market})$&Number&$p_{active}$\cr
			\midrule
			---  &---  &---  &--- &0.1 &0 &20000 &0.5\cr
			0.08 &0.04 &2000$*ratio$ &0.5 &0.08 &0.04 &2000 &0.5\cr
			0.04 &0.02 &2000$*ratio$ &0.5 &0.04 &0.02 &2000 &0.5\cr
			0.02 &0.01 &2000$*ratio$ &0.5 &0.02 &0.01 &2000 &0.5\cr
			-0.02 &0.01 &2000$*ratio$ &0.5 &-0.02 &0.01 &2000 &0.5\cr
			-0.04 &0.02 &2000$*ratio$ &0.5 &-0.04 &0.02 &2000 &0.5\cr
			-0.08 &0.04 &2000$*ratio$ &0.5 &-0.08 &0.04 &2000 &0.5\cr
			---  &---  &---  &--- &-0.1 &0 &20000 &0.5\cr
			\bottomrule
		\end{tabular}
	\end{threeparttable}
\end{table}

\section{Simulation results and preliminary analysis}
\label{simulations}
We simulate the artificial financial markets with the characteristics of pyramid schemes and obtain the rate of returns of the main fund under different cases. Through the results, we can preliminarily reveal the relationships between the rate of return of the main fund and the proportion of the trend investors in all the small investors, the small investors' parameters of taking profit and stopping loss, the order size of the main fund and the strategies adopted by the main fund.
In our simulations, the main fund participates in trading through the way of first buying and then selling, which is not fundamentally different from the situation of selling first and then buying back, this is because the parameters $r_{market}$ and numbers of small investors set in our agent model are symmetrical.

\subsection{Cases of the main fund adopting single strategy}
\label{simulation_single}
In this subsection, we assume that the main fund only adopt one strategy, which is, buying a certain amount of securities through a single order in the first trading period and selling all the securities also through one order in the next trading period. In the simulation, we also assume that all the small investors' parameters of taking profit and stopping loss different, and we consider four different combinations of taking profit parameter $r_{profit}$ and stopping loss parameter $r_{loss}$.

\begin{figure}[!htb]
	\centering
	\includegraphics[width=1\textwidth]{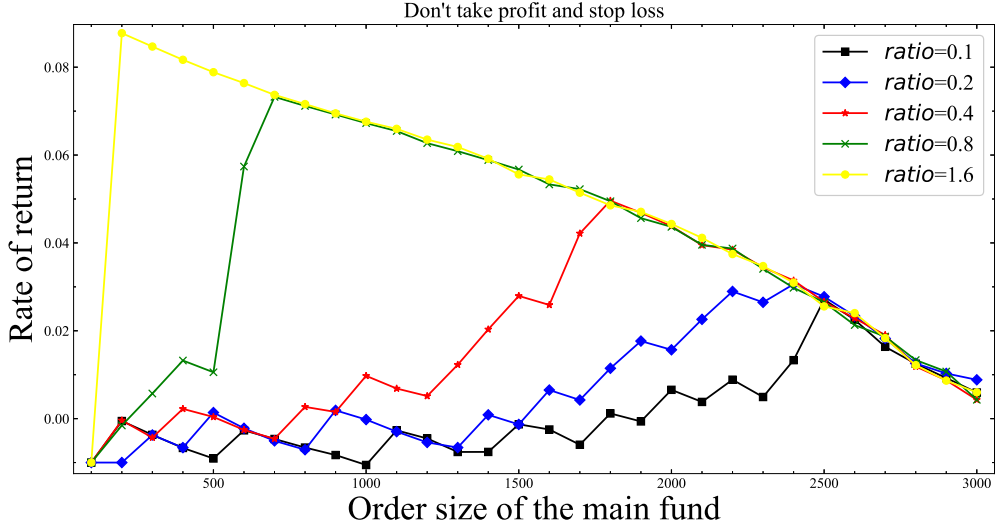}
	\caption{The relationship between the order size of the main fund and its return, when the small investors don't take profit and stop loss.}
	\label{result_no}
\end{figure}

\begin{figure}[!htb]
	\centering
	\includegraphics[width=1\textwidth]{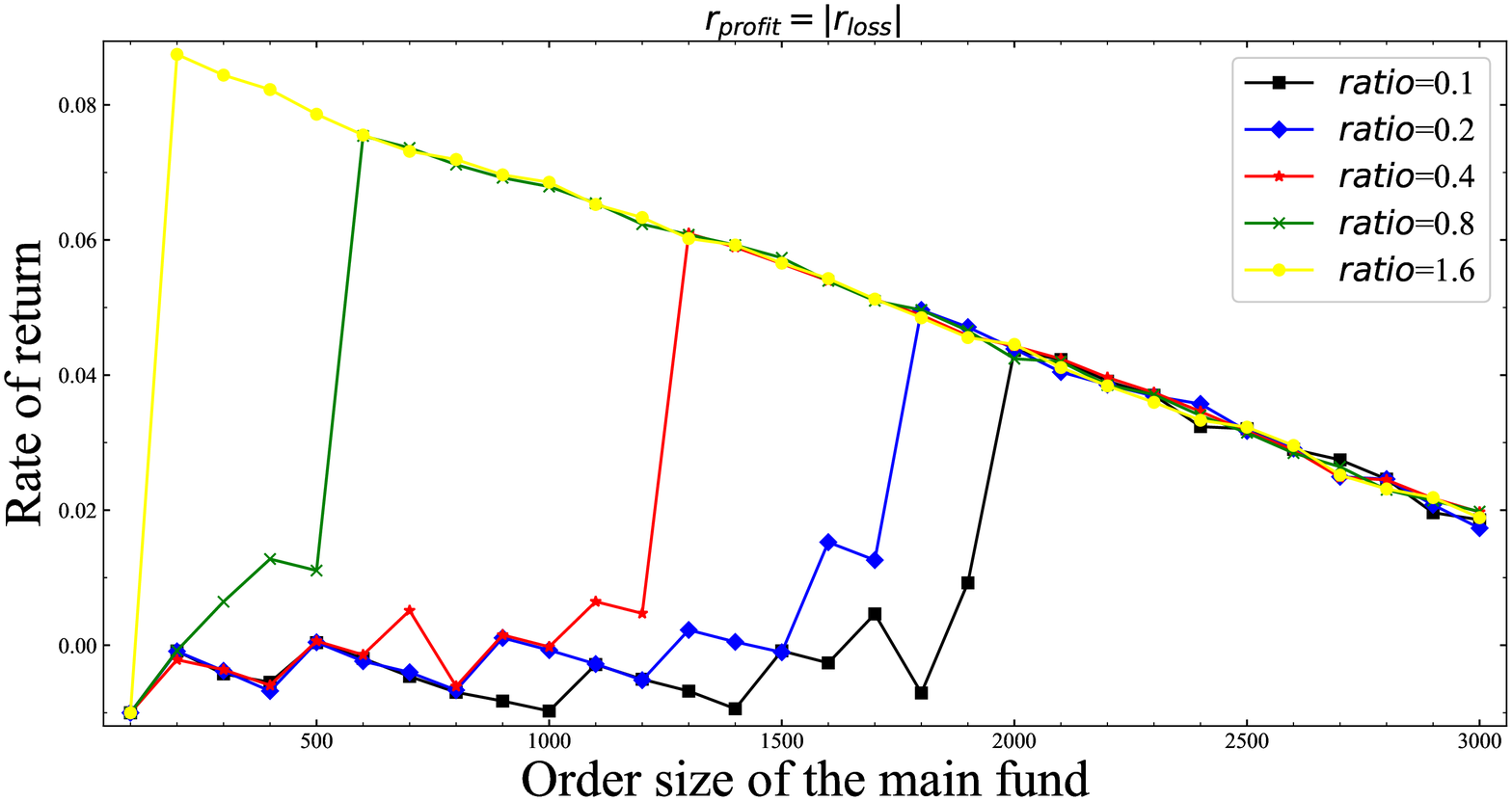}
	\caption{The relationship between the order size of the main fund and its rate of return, when the parameters of taking profit and stopping loss of the small investors satisfy $r_{profit}$ = $|r_{loss}|$, where $r_{profit}$ obeys the uniform distribution $U(0.02,0.08)$.}
	\label{result_s_e}
\end{figure}

\begin{figure}[!htb]
	\centering
	\includegraphics[width=1\textwidth]{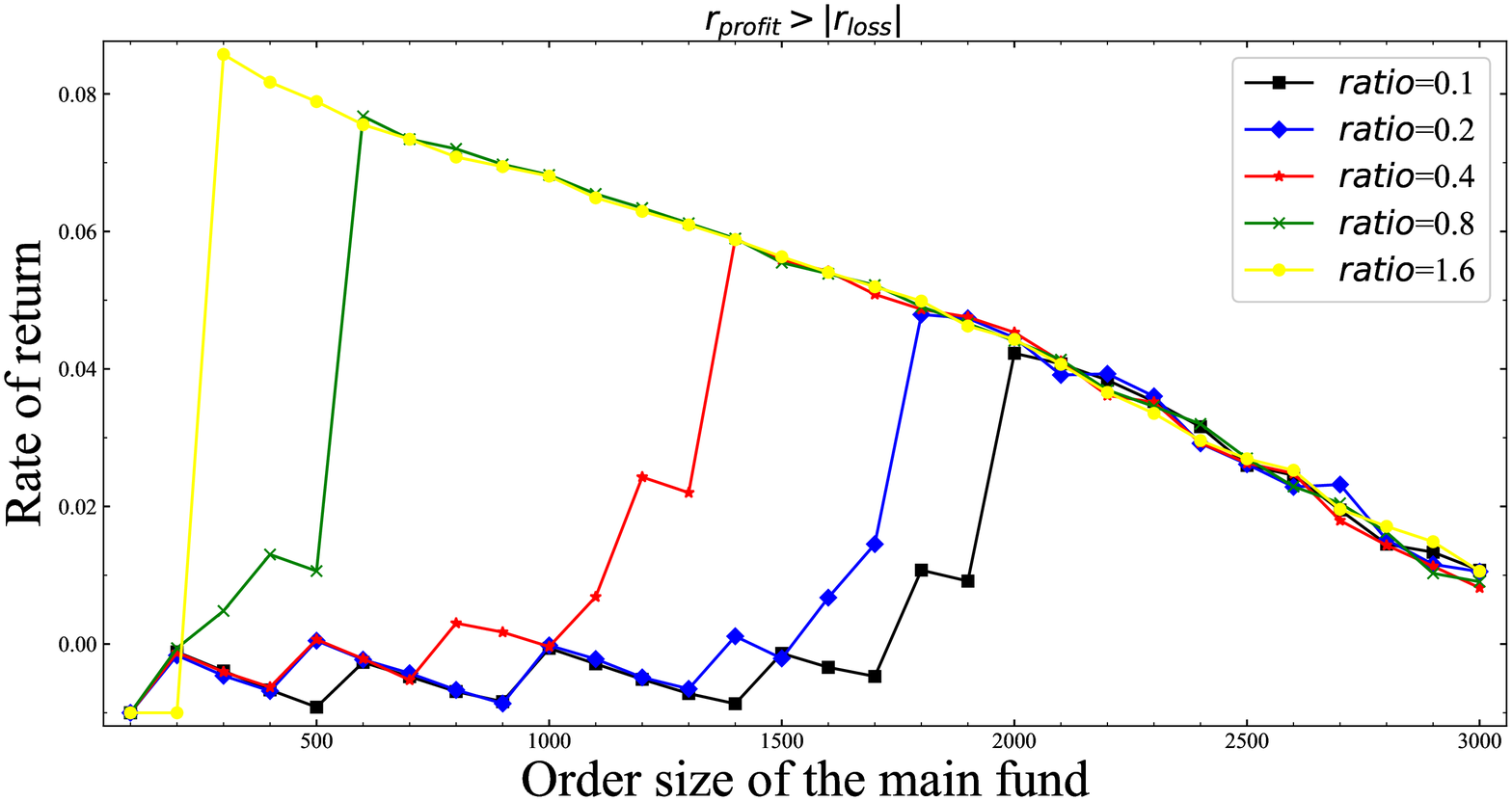}
	\caption{The relationship between the order size of the main fund and its rate of return, when the parameters of taking profit and stopping loss of the small investors satisfy $r_{profit}$ $>$ $|r_{loss}|$, where $|r_{loss}|$ obeys the uniform distribution $U(0.02,0.08)$ and $r_{profit}$ obeys the uniform distribution $U(|r_{loss}|,0.08)$.}
	\label{result_s_g}
\end{figure}

\begin{figure}[!htb]
	\centering
	\includegraphics[width=1\textwidth]{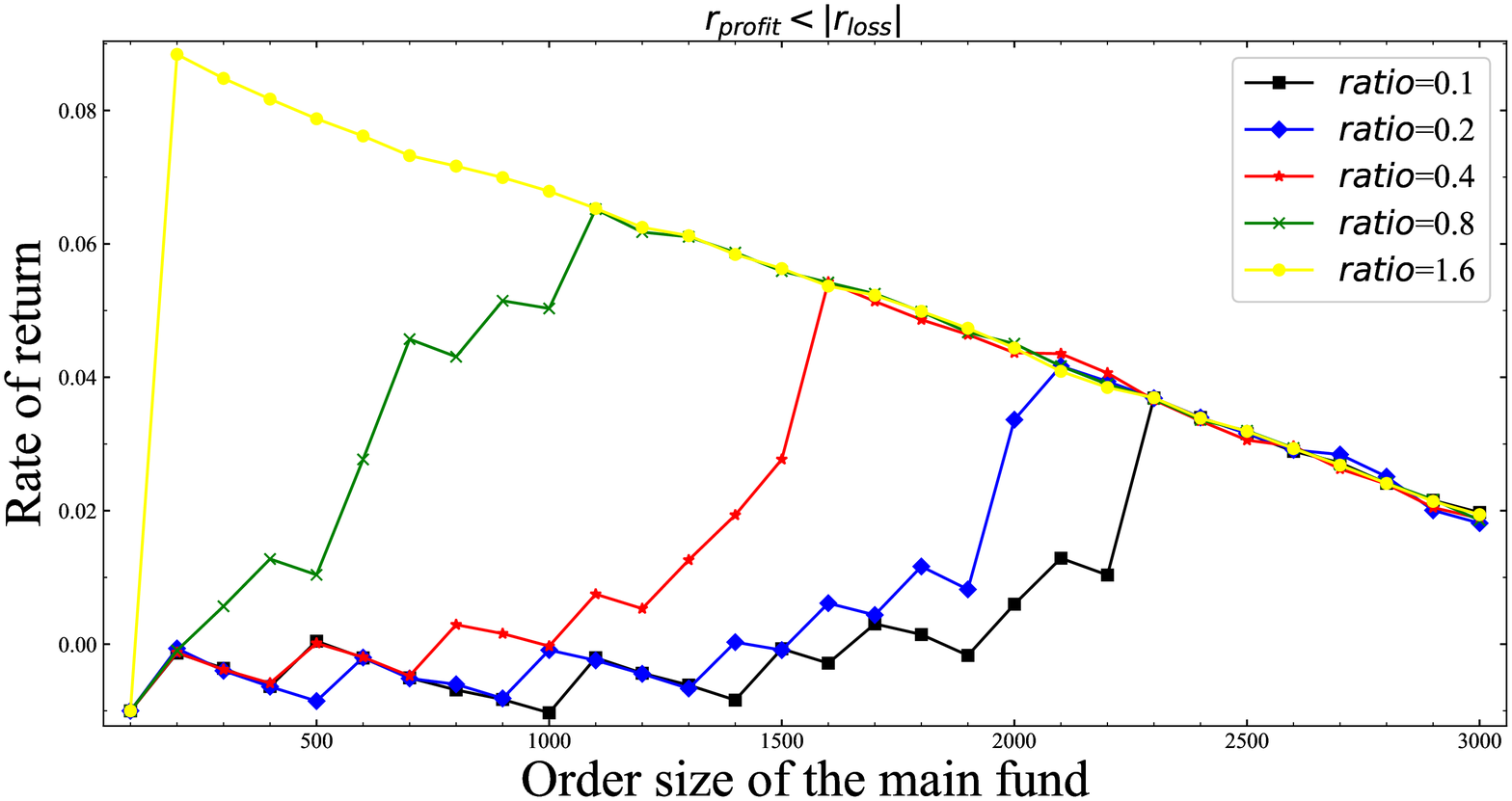}
	\caption{The relationship between the order size of the main fund and its rate of return, when the parameters of taking profit and stopping loss of the small investors satisfy $r_{profit}$ $<$ $|r_{loss}|$, where $r_{profit}$ obeys the uniform distribution $U(0.02,0.08)$ and $|r_{loss}|$ obeys the uniform distribution $U(r_{profit},0.08)$.}
	\label{result_s_l}
\end{figure}

The first case is the small investors don't take profit and stop loss, we obtain the relationship between the order size of the main fund and its return, as shown in Fig.~\ref{result_no}. Fig.~\ref{result_s_e}, Fig.~\ref{result_s_g} and Fig.~\ref{result_s_l} shows shows the relationship between the order size of the main fund and its rate of return under different parameter combinations of small investors taking profit and stopping loss respectively. The parameter combinations of small investors taking profit and stopping loss from Fig.~\ref{result_s_e} to Fig.~\ref{result_s_l} respectively are: $r_{profit}$ = $|r_{loss}|$, where $r_{profit}$ obeys the uniform distribution $U(0.02,0.08)$; $r_{profit}$ $>$ $|r_{loss}|$, where $|r_{loss}|$ obeys the uniform distribution $U(0.02,0.08)$ and $r_{profit}$ obeys the uniform distribution $U(|r_{loss}|,0.08)$; $r_{profit}$ $<$ $|r_{loss}|$, where $r_{profit}$ obeys the uniform distribution $U(0.02,0.08)$ and $|r_{loss}|$ obeys the uniform distribution $U(r_{profit},0.08)$. In addition, we consider different values of $ratio$ which are 0.1, 0.2, 0.4, 0.8 and 1.6 in these simulations.

From these figures, we can see that when the order size of the main find reaches a critical value, its yield will suddenly increase, which is due to the distribution of $r_{market}$ of the small investors is sparse when the parameters $r_{market}$ are at larger values, and when the yield reaches the maximum, it will decrease with the increase of the order size. And we also find that the higher the proportion of trend investors, the smaller the order size required by the main fund to achieve the maximum yield. Besides above, we also find that different combinations of stopping loss and taking profit parameters of small investors have an impact on the yield of the main fund.

In our artificial financial market, the return of the main fund comes from the trend investors being triggered to push up the price, and the contrarian investors accepting and buying the securities at a higher price in the next trading period.
The reason why the rate of return of the main fund increases first with the order size is that the larger order size can trigger more trend investors which help the main fund to push the price of the security higher, which increases the yield of the main fund.
The reason why the rate of return of the main fund increases quickly first and then decreases slowly is that in the parameter setting of our agent model, only 2000 investors with $r_{market}$ parameter distributed around 0.02, 0.04 and 0.08, while there are 20000 contrarian investors with parameters $r_{market}$ of 0.1, which can't be swallowed by the main fund in our model, so with the increase of the order size, the yield of the main fund increases quickly when the rise of the security's price is below 0.1, and decreases slowly when the rise of the security's price is close to 0.1. In section \ref{Theoretical_analysis}, we will make an in-depth theoretical analysis of the above conclusions.

\subsection{Cases of the main fund adopting different strategies}\label{simulation_different}
Next, we consider the cases that the main fund adopts the strategy of buying or selling  a certain amount of securities in batches in multiple trading periods. Specifically, we assume that the main fund buy a certain amount of securities in $D_{buy}$ trading periods and sell them in next $D_{sell}$ trading periods, where $D_{buy}\in[1,5]$ and $D_{sell}\in[1,5]$. In the actual financial market, the probability of no change in investor structure in the short term is higher than that in the long term, so $D_{buy}$ and $D_{sell}$ should not be taken as larger values in our model. In addition, we also assume that in the process of buying or selling securities, the main fund only places one market price order in each trading period, and the sizes of the market price orders in different trading periods are the same. Thus there are $5\times 5 = 25$ cases in total. In the simulation, the total amount of the securities that the main fund buy is set as 2000, and $ratio$ is set as $0.4$. When setting these two parameters, we refer to the simulation results in the previous subsection. Same as the previous subsection, we also assume that all the small investors' parameters of taking profit and stopping loss are different, and we consider four different situations.

\begin{figure}[!htb]
	\centering
	\includegraphics[width=1\textwidth]{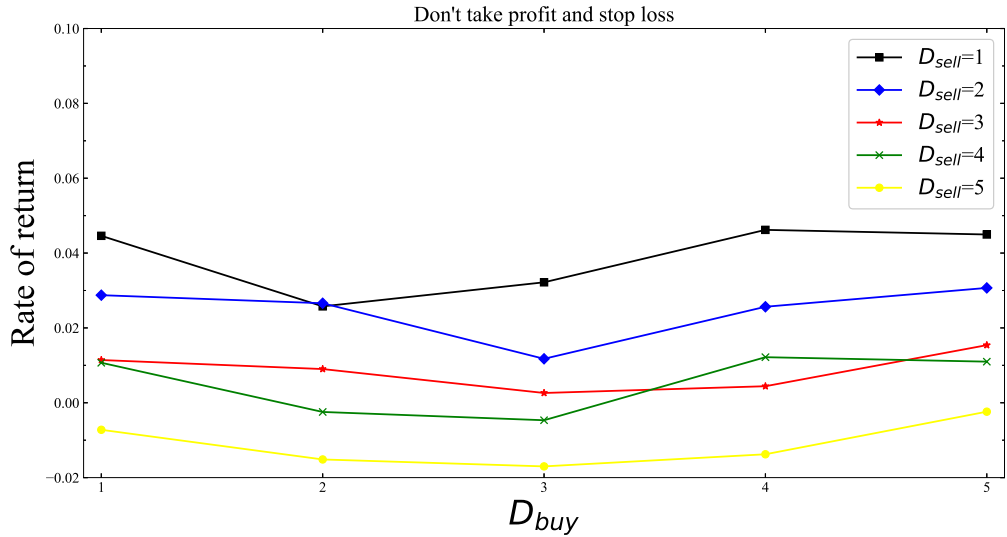}
	\caption{The relationship between the number of buying and selling periods of the main fund and its rate of return when the small investors don't take profit and stop loss.}\label{result_d_no}
\end{figure}

\begin{figure}[!htb]
	\centering
	\includegraphics[width=1\textwidth]{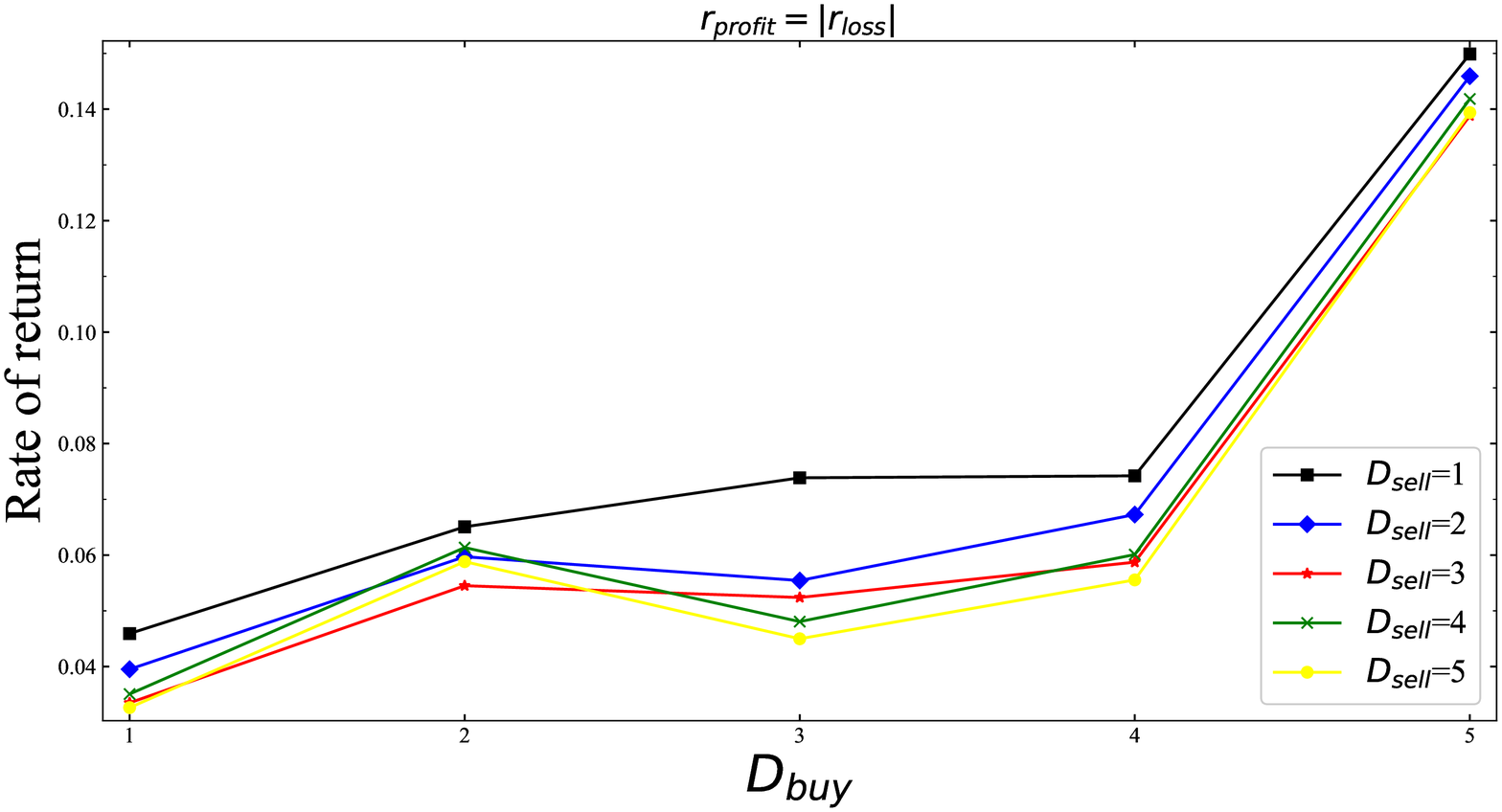}
	\caption{The relationship between the number of buying and selling periods of the main fund and its rate of return when the parameters of taking profit and stopping loss of the small investors satisfy $r_{profit}$ = $|r_{loss}|$, where $r_{profit}$ obeys the uniform distribution $U(0.02,0.08)$.}\label{result_d_e}
\end{figure}

\begin{figure}[!htb]
	\centering
	\includegraphics[width=1\textwidth]{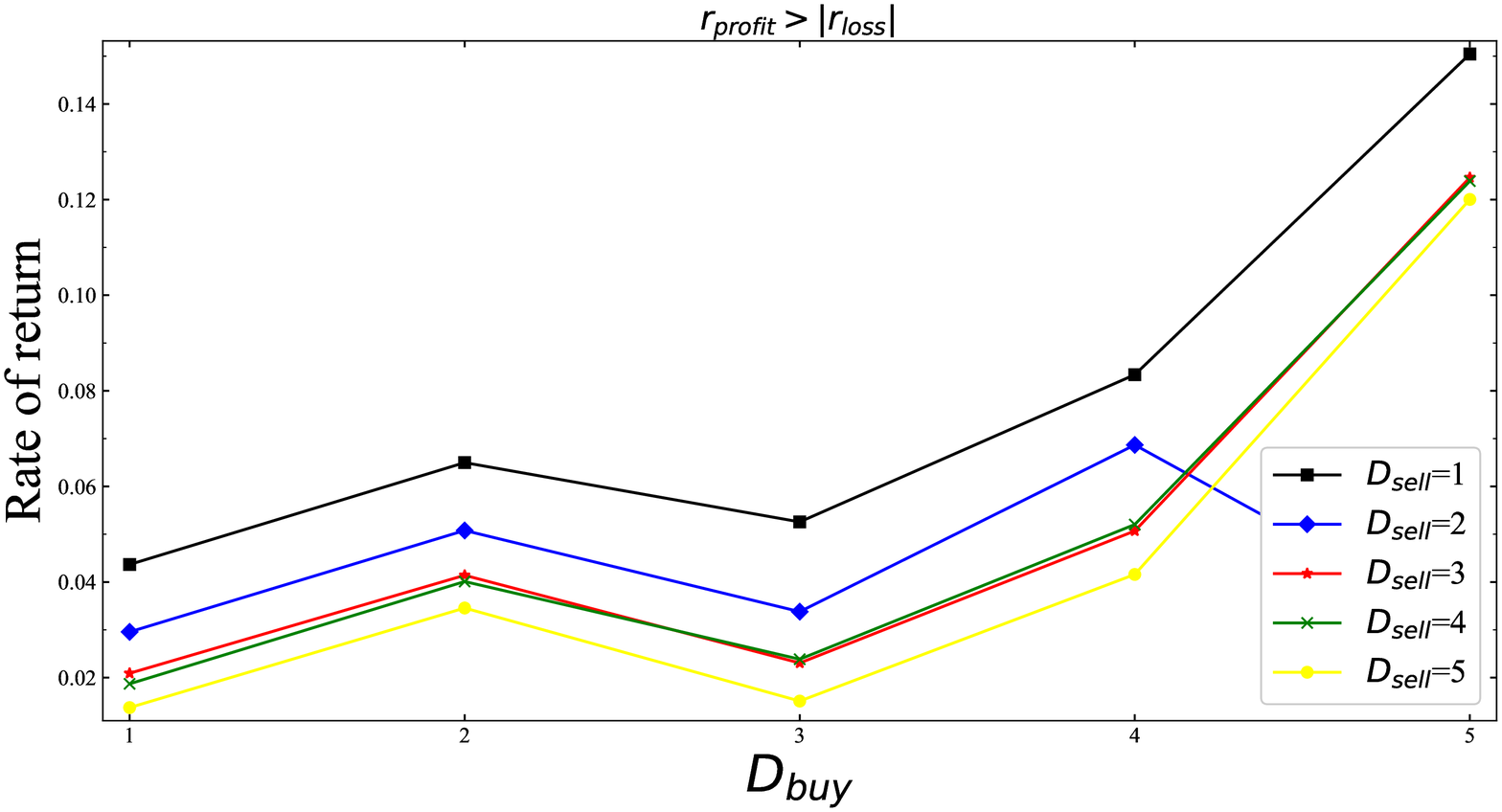}
	\caption{The relationship between the number of buying and selling periods of the main fund and its rate of return when the parameters of taking profit and stopping loss of the small investors satisfy $r_{profit}$ $>$ $|r_{loss}|$, where $|r_{loss}|$ obeys the uniform distribution $U(0.02,0.08)$ and $r_{profit}$ obeys the uniform distribution $U(|r_{loss}|,0.08)$.}\label{result_d_g}
\end{figure}

\begin{figure}[!htb]
	\centering
	\includegraphics[width=1\textwidth]{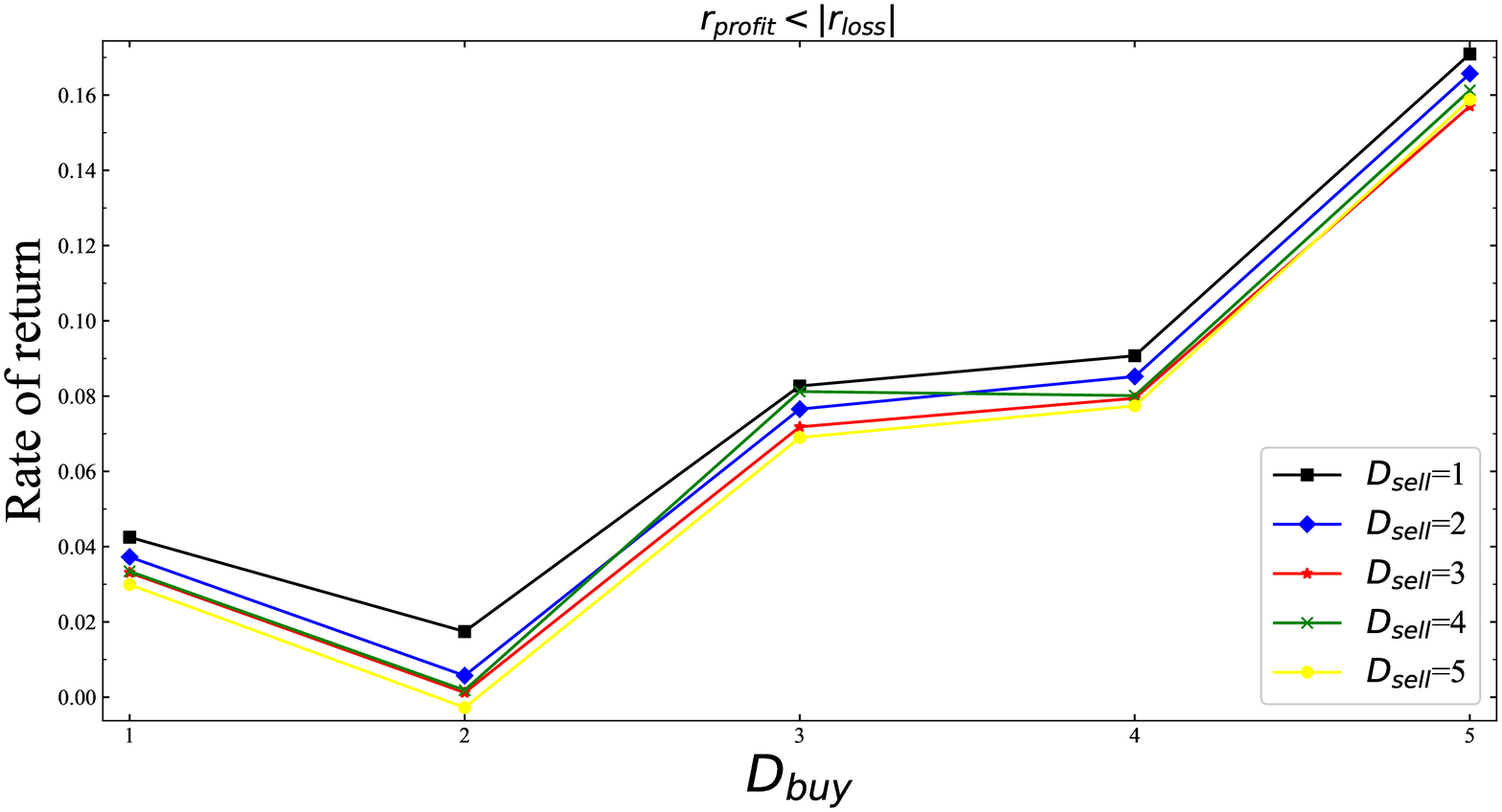}
	\caption{The relationship between the number of buying and selling periods of the main fund and its rate of return when the parameters of taking profit and stopping loss of the small investors satisfy $r_{profit}$ $<$ $|r_{loss}|$, where $r_{profit}$ obeys the uniform distribution $U(0.02,0.08)$ and $|r_{loss}|$ obeys the uniform distribution $U(r_{profit},0.08)$.}\label{result_d_l}
\end{figure}

The first situation is that the small investors don't take profit and stop loss, we obtain the relationship between the number of buying and selling periods of the main fund and its rate of return, as shown in Fig.~\ref{result_d_no}. Fig.~\ref{result_d_e}, Fig.~\ref{result_d_g} and Fig.~\ref{result_d_l} shows the relationship between the number of buying and selling trading periods of the main fund and its rate of return under different parameter combinations of small investors taking profit and stopping loss respectively.  The parameter combinations of small investors taking profit and stopping loss we consider also are $r_{profit}$ = $|r_{loss}|$, where $r_{profit}$ obeys the uniform distribution $U(0.02,0.08)$; $r_{profit}$ $>$ $|r_{loss}|$, where $|r_{loss}|$ obeys the uniform distribution $U(0.02,0.08)$ and $r_{profit}$ obeys the uniform distribution $U(|r_{loss}|,0.08)$; $r_{profit}$ $<$ $|r_{loss}|$, where $r_{profit}$ obeys the uniform distribution $U(0.02,0.08)$ and $|r_{loss}|$ obeys the uniform distribution $U(r_{profit},0.08)$. The above different combinations correspond to the simulation results in Fig.~\ref{result_d_e} to Fig.~\ref{result_d_l} respectively.

From these figures, we can find that under different investor structures, the strategies and corresponding yields of the main fund are different. When the small investors don't take profit and stop loss, the yield curve of the main fund is basically the lowest. Moreover, there are some trends in these figures, that is, the more trading periods of the main fund buying securities, the higher the yield, and the fewer trading periods of the main fund selling securities, the higher the rate of return. Through more careful observation, we find that the effect of changes in the trading periods' number of buying on the yield seems greater than that of the trading periods' number of selling. In addition, we also find that if the investors do not take profit and stop loss, the yield of the main fund is significantly smaller than other situations.

This shows in the case of unchanged investor structure, the main fund tend to increase the number of buying periods and reduce the selling periods, which can explain why the real stock market always rises slowly and falls quickly. In the next section, we will make a more in-depth theoretical analysis of the conclusion.

\section{Theoretical analysis}
\label{Theoretical_analysis}
To further understand the underlying mechanism of the proposed agent-based model and avoid illusive results caused by the model setting, we conduct a theoretical analysis on the model as follows.
In the section, we analyze in detail the effects of the order size $N_{mf}$ of the main fund, the parameter $ratio$ between the trend investors and the contrarian
investors, the stop-loss parameters $r_{loss}$, take-profit parameters $r_{profit}$ and the activation probability $p_{active}$ of the small investors on the yield of the main fund when the main fund adopt a single strategy.
For the cases of that the main fund adopts the strategies of buying or selling a certain equal amount of securities in batches in multiple trading periods, we analyze the impact of different numbers of buying or selling trading periods on the main fund yield.
\subsection{Cases of the main fund adopting single strategy}
\label{theoretical_single}
For the cases of the main fund adopting single strategy described in section 3.1, we firstly arrange the parameters $r_{market}$ of the contrarian investors and trend investors into two series
$\{R_M^{C-},$ $\cdots,$ $R_2^{C-},$ $R_1^{C-},$ $R_1^{C+},$ $R_2^{C+},$ $\cdots,$  $R_N^{C+}\}$ and
$\{R_P^{T-},$ $\cdots,$ $R_2^{T-},$ $R_1^{T-},$ $R_1^{T+},$ $R_2^{T+},$ $\cdots,$  $R_Q^{T+}\}$ respectively,
where $M$, $N$, $P$ and $Q$ denote the numbers of the contrarian investors and trend investors who have negative and positive $r_{market}$, $R_i^{C-}$ $\textless$ $0$, $R_j^{C+}$ $\textgreater$ $0$, $R_k^{T-}$ $\textless$ $0$ and $R_l^{T+}$ $\textgreater$ $0$ where $i$, $j$, $k$ and $l$ are positive integers in $[1,M]$, $[1,N]$, $[1,P]$ and $[1,Q]$ respectively,
$R_M^{C-}$ $\leq$ $\cdots$ $\leq$ $R_2^{C-}$ $\leq$ $R_1^{C-}$ $\textless$ $R_1^{C+}$ $\leq$ $R_2^{C+}$ $\leq$ $\cdots$ $\leq$ $R_N^{C+}$ and
$R_P^{T-}$ $\leq$ $\cdots$ $\leq$ $R_2^{T-}$ $\leq$ $R_1^{T-}$ $\textless$ $R_1^{T+}$ $\leq$ $R_2^{T+}$ $\leq$ $\cdots$ $\leq$ $R_Q^{T+}$.
Since we set an activation probability $p_{active}$ for each small investor, we assume that the sequences of parameters $r_{market}$ of the small investors activated in trading are $\{r_m^{C-},$ $\cdots,$ $r_2^{C-},$ $r_1^{C-},$ $r_1^{C+},$ $r_2^{C+},$ $\cdots,$  $r_n^{C+}\}$ and
$\{r_p^{T-},$ $\cdots,$ $r_2^{T-},$ $r_1^{T-},$ $r_1^{T+},$ $r_2^{T+},$ $\cdots,$  $r_q^{T+}\}$,
where $m$, $n$, $p$ and $q$ denote the numbers of the activated contrarian investors and trend investors with negative and positive $r_{market}$.
According to the hypotheses of our model, after the transaction of the market buy order with the scale of $N_{mf}$ issued by the main fund ($N_{mf}$ is an integer and less than $n$), the price of the security goes up to $(1+r_{N_{mf}}^{C+})P_0$ where $P_0$ is the initial price of the security, and the cost of the main fund is given by
\begin{equation}
M_{buy} = \sum_{i=1}^{N_{mf}}(1+r_{i}^{C+})P_0.
\end{equation}
The price rise may trigger some trend investors, and the price rise caused by these trend investors' buying may trigger new trend investors. The numbers of the trend investors triggered for the first time $N_{t}^1$, the second time $N_{t}^2$, ..., and the $i$-th time $N_{t}^i$ are given by
\begin{equation}
\begin{aligned}
&N_{t}^1 = \sum_{i=1}^{q} \mathbb{I}(r_{i}^{T+}\leq r_{N_{mf}}^{C+}),
N_{t}^2 = \sum_{i>N_{t}^1}^{q} \mathbb{I}(r_{i}^{T+}\leq r_{N_{mf}+N_{t}^1}^{C+}),\cdots,\\
&N_{t}^i = \sum_{i>N_{t}^{i-1}}^{q} \mathbb{I}(r_{i}^{T+}\leq r_{N_{mf}+N_{t}^1+N_{t}^2+\cdots+N_{t}^{i-1}}^{C+}),&\\
\end{aligned}
\end{equation}
where $\mathbb{I}(\textbf{x})$ is the indicator function, when $\textbf{x}$ is true $\mathbb{I}(\textbf{x})$ is equal to 1, otherwise it is 0. We assume that this process lasts a total of $x$ times, and define $N_t^{X}$ $=$ $\sum^{x}_{i=1}N_t^i$, then the security price at the end of the first trading period is $P(1)$ $=$ $(1+r_{N_{mf}+N_t^{X}}^{C+})P_0$.
It should be noted that in our model setting, the contrarian investors are sufficient so that the expression $N_{mf}+N_t^{X}$ $<$ $n$ is satisfied. In addition, in the above analysis, the small investors don't take profit and stop loss. In the second trading period, the main fund sell the securities they have bought, and according to a similar analysis, the security price at the end of the second trading period is $P(2)$ $=$ $(1+r_{N_{mf}+N_t^{Y}}^{C-})P(1)$, where $N_t^Y$ is the number of trend investors triggered in this trading period. The money $M_{sell}$ of the main fund obtaining from selling the securities is given by
\begin{equation}
M_{sell} = \sum_{i=1}^{N_{mf}}(1+r_{i}^{C-})P(1),
\end{equation}
then the rate of return of the main fund $r_{mf}$ is
\begin{equation}
\begin{aligned}
r_{mf}&=\frac{M_{sell}}{M_{buy}}-1
=\frac{(1+r_{N_{mf}+N_{t}^X}^{C+})\sum_{i=1}^{N_{mf}}(1+r_{i}^{C-})}{\sum_{i=1}^{N_{mf}}(1+r_{i}^{C+})}-1&\\
&=(1+r_{N_{mf}+N_{t}^X}^{C+})\frac{1+\sum_{i=1}^{N_{mf}}r_{i}^{C-}/N_{mf}}{1+\sum_{i=1}^{N_{mf}}r_{i}^{C+}/N_{mf}}-1.
\end{aligned}
\end{equation}
The above expression of $r_{mf}$ can explain our simulation results presented in Fig.~1. The left term $(1+r_{N_{mf}+N_{t}^X}^{C+})$ increases with the increase of $N_{mf}$ until it reaches the maximum (the reason for the existence the maximum is that there are enough contrarian investors when the parameters $r_{market} $ are close to 0.1), and the right term $(1+\sum_{i=1}^{N_{mf}}r_{i}^{C-}/N_{mf})/(1+\sum_{i=1}^{N_{mf}}r_{i}^{C+}/N_{mf})$ decreases with the increase of $N_{mf}$, where $\sum_{i=1}^{N_{mf}}r_{i}^{C-}/N_{mf}$ and $\sum_{i=1}^{N_{mf}}r_{i}^{C+}/N_{mf}$ are actually the mean value of the sets $\{r_i^{C-}$ $|$ $i\in[1, N_{mf}]\cap N^+\}$ and $\{r_i^{C+}$ $|$ $i\in[1, N_{mf}]\cap N^+\}$, and they reflect the increased average cost of the main fund due to the price rise caused by its own selling and buying. Because $\{R_M^{C-},$ $\cdots,$ $R_2^{C-},$ $R_1^{C-},$ $R_1^{C+},$ $R_2^{C+},$ $\cdots,$ $R_N^{C+}\}$ in our model is symmetric and the activation probability of each small investor is equal, so the transaction situations of the main fund buying and selling are almost the same, and it is easy to find that when $r_{N_{mf}+N_{t}^X}^{C+}$ does not reach the maximum, the left term increases faster than the right term with the increase of $N_{mf}$, therefore, $r_{mf}$ increases with the increase of $N_{mf}$ and decreases when $r_{N_{mf}+N_{t}^X}^{C+}$ reaches the maximum, which is consistent with the simulation results presented in Fig.~\ref{result_no}.
In addition, the larger the ratio parameter $ratio$ between the trend investors and the contrarian investors, the larger the number $N_t^X$ of the triggered trend investors, and $r_{mf}$ will reach the maximum faster, which is also consistent with the simulation results presented in Fig.~\ref{result_no}.
Here we must point out that the reason why the main fund gains profits is that the contrarian investors whose parameter $r_{market}$ is less than 0 accept the new benchmark price $P(1)$ in the second trading period.

Next, we analyze the effect of small investors' taking profit and stopping loss parameters on the yield of the main fund. In the first trading period, the contrarian investors making deals with the main fund place the following stop-loss orders $\{-r_{loss}(1+r_i^{C+}){P_0}/{P(D-1)}\}$ and take-profit orders $\{-r_{return}(1+r_i^{C+}){P_0}/{P(D-1)}\}$, where $D$ is the serial number of the trading period and $P(0)=P_0$. The take-profit orders will be added to the new sequence $\{r_m^{C-},$ $\cdots,$ $r_2^{C-},$ $r_{1}^{C-},$ $r_{N_{mf}+1}^{C+},$ $r_{N_{mf}+2}^{C+},$ $\cdots,$  $r_n^{C+}\}$ after trading with the main fund, and the stop-loss orders will be added to the sequence $\{r_p^{T-},$ $\cdots,$ $r_2^{T-},$ $r_1^{T-},$ $r_1^{T+},$ $r_2^{T+},$ $\cdots,$  $r_q^{T+}\}$. We find that for the main fund, the smaller the absolute values of the stop-loss parameters of the contrarian investors, the easier it is for the contrarian investors to stop loss and help the main fund to push up the security price; the smaller the take-profit parameters of the contrarian investors are, the higher the prices the contrarian investors buy back the securities from the main fund in the trading periods when the main fund is selling. Therefore, the smaller the stop-loss and take-profit parameters of reverse investors, the higher the yield of the main fund. Similar analysis of the trend investors shows that the larger the absolute values of the taking profit and stopping loss parameters, the higher the yield of the main fund.

\begin{figure}[!htb]
	\centering
	\includegraphics[width=1\textwidth]{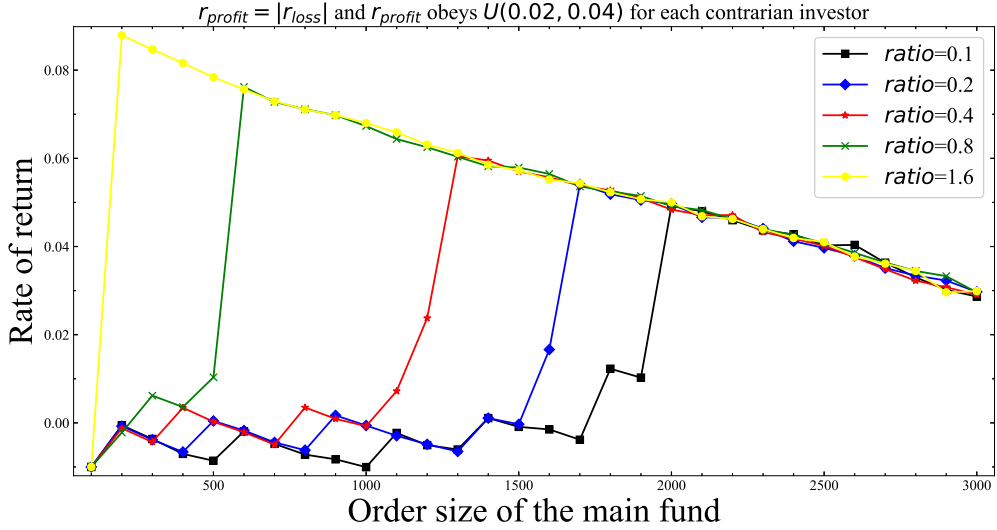}
	\caption{The relationship between the order size of the main fund and its rate of return, when the parameters of taking profit and stopping loss of the contrarian investors satisfy $r_{profit}$ = $|r_{loss}|$, where $r_{profit}$ obeys the uniform distribution $U(0.02,0.04)$, and the contrarian investors don't take profit and stop loss.}\label{result_s_c2}
\end{figure}

\begin{figure}[!htb]
	\centering
	\includegraphics[width=1\textwidth]{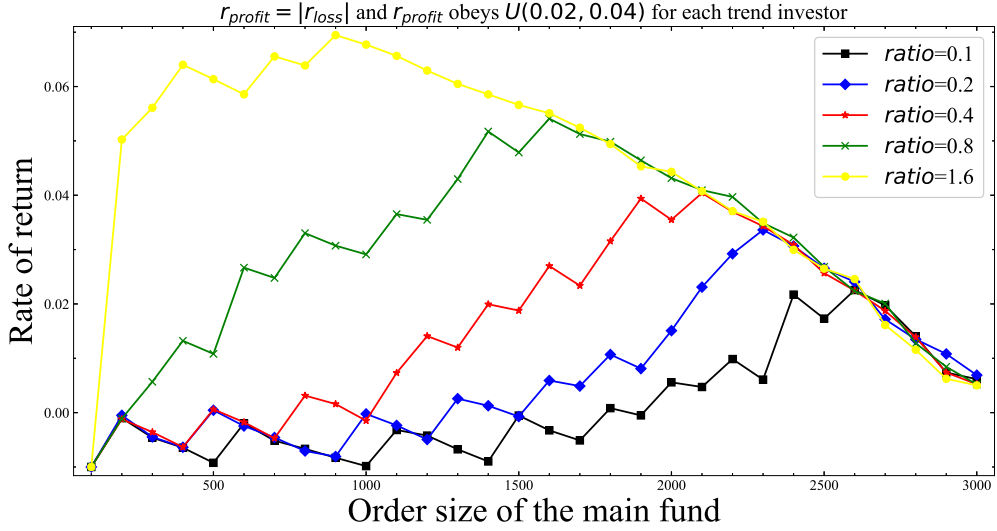}
	\caption{The relationship between the order size of the main fund and its rate of return, when the parameters of taking profit and stopping loss of the trend investors satisfy $r_{profit}$ = $|r_{loss}|$, where $r_{profit}$ obeys the uniform distribution $U(0.02,0.04)$, and the trend investors don't take profit and stop loss.}\label{result_s_t2}
\end{figure}

We design simulation experiments to verify our conclusion. We set the absolute values of stop-loss and take-profit parameters of the contrarian investors and trend investors to random values obeying the uniform distribution $U(0.02,0.04)$ respectively without setting take-profit parameters and stop-loss parameters of the trend investors or contrarian investors (no take-profit parameters and stop-loss parameters are equivalent to the infinity of the absolute values of them). Comparing our simulation results in Fig.~\ref{result_s_c2} and Fig.~\ref{result_s_t2} with the results in Fig.~\ref{result_no}, we find that the returns in Fig.~\ref{result_s_c2} are generally higher than those in Fig.~\ref{result_no} and the returns in Fig.~\ref{result_s_t2} are generally lower than those in Fig.~\ref{result_no}, which are consistent with our analysis: the smaller the absolute values of the take-profit and stop-loss parameters of the contrarian investors, the higher the yield of the main fund, while the situation of the trend investors is just the opposite.

At the end of this subsection, we analyze the influence of the activation probability $p_{active}$ on $r_{mf}$. In fact, the effect of lower $p_{active}$ is equivalent to that of larger $N_{mf}$, so within a certain parameter range, the value of $r_{mf}$ increases with the decrease of $p_{active}$ and then decreases when $r_{mf}$ reaches the maximum.

\subsection{Cases of the main fund adopting different strategies}
For the cases that the main fund adopts the strategies of buying or selling a certain equal amount of securities in batches in multiple trading periods described in section 3.2, we firstly assume $N_{mf}$ $=$ $n_{mf}^b D_{buy}$ =$n_{mf}^s D_{sell}$, where $n_{mf}^b$ and $D_{buy}$ are the number of securities bought per trading period by the main fund and the total number of trading periods of buying, $n_{mf}^s$ and $D_{sell}$ are the corresponding quantities in the trading periods of selling. In this subsection, we continue to use the notations defined in section~\ref{theoretical_single}, but note that $r_i^{C+}$, $r_i^{C-}$ and $N_t^X$ are not the same in different trading periods, we denote them as $r_i^{C+}(k)$,  $r_i^{C-}(k)$ and $N_t^X(k)$ respectively, where $k$ is the serial number of the trading period of buying or selling. As the section section~\ref{theoretical_single} has considered the effect of $r_{profit}$ and $r_{loss}$ on the main fund yield, we will not consider it in this subsection. The total cost $M^D_{buy}$ is the sum of the cost $M_{buy}$ in each trading period, which is given by
\begin{equation}
\begin{aligned}
M^D_{buy} = \sum_{j=1}^{D_{buy}}M_{buy}(j)
= \sum_{j=1}^{D_{buy}}\sum_{i=1}^{n_{mf}^b}[1+r^{C+}_i(j)]\prod \limits_{k=1}^{j}[1+f_b(k-1)]P_0
\end{aligned}
\end{equation}
where the function $f_b(k)$ $=$ $r^{C+}_{n_{mf}^b+N_t^{X}(k)}(k)$, $k$ is the serial number of the trading period of buying and we define $f_b(0)=0$.
The total money $M_{sell}$ of the main fund obtaining from selling the securities in multiple trading periods is given by
\begin{equation}
\begin{aligned}
M^D_{sell} = \sum_{j=1}^{D_{sell}}M_{sell}(j)
= \sum_{j=1}^{D_{sell}}\sum_{i=1}^{n_{mf}^s}[1+r^{C-}_i(j)]\prod \limits_{k=1}^{j}[1+f_s(k-1)]P_0^s
\end{aligned}
\end{equation}
where the function $f_s(k)$ $=$ $r^{C-}_{n_{mf}^s+N_t^{Y}(k)}(k)$, $k$ is the serial number of the trading period of selling and we define $f_s(0)=0$, $N_t^{Y}(k)$ is the corresponding quantity of $N_t^{X}(k)$ in the trading periods of selling, $P_0^s$ is the price of the security at the end of the last trading period of buying and $P_0^s=\prod \limits_{l=1}^{D_{buy}}[1+f_b(l)]P_0$. Then
\begin{equation}
\begin{aligned}
r_{mf} &=\frac{M_{sell}^D}{M_{buy}^D}-1
=\prod \limits_{l=1}^{D_{buy}}[1+f_b(l)]\frac{\sum_{j=1}^{D_{sell}}\sum_{i=1}^{n_{mf}^s}[1+r^{C-}_i(j)]\prod \limits_{k=1}^{j}[1+f_s(k-1)]}{\sum_{j=1}^{D_{buy}}\sum_{i=1}^{n_{mf}^b}[1+r^{C+}_i(j)]\prod \limits_{k=1}^{j}[1+f_b(k-1)]}-1\\
&\approx [1+\sum_{j=1}^{D_{buy}}f_b(j)]\frac{1+\sum^{D_{sell}}_{j=1}[\sum^{n^s_{mf}}_{i=1}r_i^{C-}(j)/n_{mf}^s+\sum^j_{k=1}f_s(k-1)]/D_{sell}}{1+\sum^{D_{buy}}_{j=1}[\sum^{n^b_{mf}}_{i=1}r_i^{C+}(j)/n_{mf}^b+\sum^j_{k=1}f_b(k-1)]/D_{buy}}-1,
\end{aligned}
\end{equation}
in the approximation, we only keep the first order quantity of $r^{C+}_i$ or $r^{C-}_i$, because in our simulations the absolute values of $r^{C+}_i$ and $r^{C-}_i$ are no more than 0.1, and the absolute values of $\sum^{D_{sell}}_{k=1}f_s(k-1)$ and $\sum^{D_{buy}}_{k=1}f_b(k-1)$ are no more than 0.4. We define $g_b(j)$=$\sum^{n^b_{mf}}_{i=1}r_i^{C+}(j)/n_{mf}^b/D_{buy}+\sum^j_{k=1}f_b(k-1)/D_{buy}$, and the term $\sum^{n^b_{mf}}_{i=1}r_i^{C+}(j)/n_{mf}^b$ in $g_b(j)$ measures the increased average cost of the main fund due to the price rise caused by its own buying in the $j$-th trading period, which is less sensitive to the change of $n_{mf}^b$ than the highest rise of the security represented by $f_b(j)$ in the same trading period. Then, we explain the change of the term $\sum^j_{k=1}f_b(k-1)/D_{buy}$ with $n_{mf}^b$ is not as sensitive as that of $f_b(j)$ with $n_{mf}^b$ as follows, which measures the increased cost due to the price rise caused by the main fund buying before the $j$-th trading periods.

The greater $j$ is, the smaller the number of the trend and contrarian investors whose parameters $r_{market}$ are greater than 0 because some of them have traded with the main fund, and the contrarian investors play a leading role, this is because in our model settings, the number of them is much larger than that of the trend investors and the distributions of trend investors and contrarian investors are the same except for their numbers. Therefore, with the increase of $j$, the main fund will be more easy to push up the security price, that is, larger $j$ leads greater increase of $f_b(j)$ when $n_{mf}^b$ increases, so $f_b(j)$ is more sensitive to the change of $n_{mf}^b$ with larger $j$, then $f_b(j)$ is more sensitive to the change of $n_{mf}^b$ than $\sum^j_{k=1}f_b(k-1)/D_{buy}$.

Based on the above analysis, the term $[1+\sum_{j=1}^{D_{buy}}f_b(j)]$ plays a leading role in the change of $r_{mf}$ when $n_{mf}^b$ changes. As $n_{mf}^b$ decreases, $D_{buy}$ will increase, and $[1+\sum_{j=1}^{D_{buy}}f_b(j)]$ will also increase according to the above analysis of $f_b(j)$. So in our model setting, the main fund increasing the number of times of buying will increase its yield. A similar analysis of $\{1+\sum^{D_{sell}}_{j=1}[\sum^{n^s_{mf}}_{i=1}r_i^{C-}(j)/n_{mf}^s+\sum^j_{k=1}f_s(k-1)]/D_{sell}\}$ shows that decreasing $D_{sell}$ will increase the yield of the main fund.

\section{Conclusion and discussion}
\label{conclusion}
We have established an agent model to study the financial phenomenon with the characteristics of pyramid schemes, and simulated this kind of artificial financial market under different investor structures. Through the simulation results and theoretical analysis, we obtain the relationships between the rate of return of the main fund and the proportion of the small trend investors in all small investors, the small investors' parameters of taking profit and stopping loss, the order size of the main fund and the strategies adopted by the main fund. Our main conclusions are: when the main fund adopts the strategy of buying a certain amount of securities through just one market order in the first trading period and selling them also through one market order in the second period, the higher the proportion of the small trend investors, the smaller the order size of the main fund to obtain the maximum yield, and after that the yield of the main fund decreases slowly with the increase of the order size; when the main fund adopts the strategies of buying and selling securities in multiple trading periods, we find that the less trading periods of selling and the more trading periods of buying, the higher the yield.

Our results are helpful to explain the financial phenomenon with the characteristics of pyramid schemes in financial markets, design trading rules for regulators and develop trading strategies for investors. Specifically, for investors, they can use our model to develop effective investment strategies after evaluating the structure of investors in the market; for financial market regulators, through our agent model and introducing parameters in real financial market, regulators can limit the maximum order size of the big investors in a certain period of time, so as to avoid the inequality of investors with different amount of money.

\section{Acknowledgement}
This research was supported by National Natural Science Foundation of China (No.~71932008, 91546201).

\bibliographystyle{ieeetr}
\bibliography{reference}
\end{document}